\documentclass[12pt]{article}
\usepackage{amsmath}
\usepackage{latexsym}
\usepackage{bbm}

\def \ATMP {{\it Adv. Theor. Math. Phys.} }

\def \NC {{\it Nuovo Cim.}}
\def \NCL {{\it Nuovo Cim. Lett.} }
\def \NP {{\it Nucl. Phys.} }

\def \PL {{\it Phys. Lett.} }
\def \PR {{\it Phys. Rev.} }

\def \PRL {{\it Phys. Rev. Lett.} }
\def \PTP {{\it Prog. Theor. Phys.} }

\def\d{\delta}

\def\m{\mu}
\def\n{\nu}

\def\s{\sigma}
\def\t{\tau}

\def\c{\chi}

\def\L{\Lambda}

\def\IZ{\mathbbm Z}

\def\and{{\rm and}}

\def\ie{{\it i.e.,} }

\begin{document}
\vspace*{-.6in} \thispagestyle{empty}
\begin{flushright}
CALT-68-2714
\end{flushright}
\baselineskip = 18pt

\vspace{1.5in} {\Large
\begin{center}
STATUS OF SUPERSTRING AND M-THEORY
\end{center}}

\begin{center}
John H. Schwarz\footnote{jhs@theory.caltech.edu}
\\
\emph{California Institute of Technology\\ Pasadena, CA  91125, USA}
\end{center}
\vspace{1in}

\begin{center}
\textbf{Abstract}
\end{center}
\begin{quotation}
\noindent The first lecture gives a colloquium-level overview of
string theory and M-theory. The second lecture surveys various
attempts to construct a viable model of particle physics. A recently
proposed approach, based on F-theory, is emphasized.
\end{quotation}

\vspace{1.0in} \centerline{Lectures presented at the Erice 2008
International School of Subnuclear Physics}

\newpage

\pagestyle{empty}

\section{Introduction}

The plan for my two lectures is as follows: the first lecture (in
section 2) will give a colloquium level overview of string theory.
The presentation will mention some of the history of the subject at
the same time that basic concepts (such as compactification of extra
dimensions, dualities, and branes) are introduced. There is a lot of
ground to cover, and it will be necessary to be somewhat sketchy.
Since the theme of this school concerns discoveries that might be
made at the LHC, it seems appropriate to emphasize string theory
approaches to particle physics model building. Therefore this is the
subject of the second lecture (in section 3). It will survey several
different approaches that have been developed, describing some of
the successes and problems of each of them. One approach that has
seen a lot of recent progress, and which looks especially promising,
will be emphasized.

For reasons that will be explained, the emphasis is on
supersymmetrical string theories (called {\it superstring
theories}), which naturally are associated with ten-dimensional
spacetime. In certain cases, the strong coupling limit gives an
eleven-dimensional theory, called {\it M-theory}, which does not
contain strings. For the student who wants to learn more, there are
many textbooks on string theory. Some of them are
\cite{Green:1987cup,Polchinski:1998rr,Zwiebach:2004tj,Becker:2007cup}.
The one by Zwiebach \cite{Zwiebach:2004tj} is addressed to advanced
undergraduates, but it is suitable for all physicists who are not
trying to become experts. The other three attempt to bring the
reader up to the state of the art at the time when they were
written. The subject is fast-moving and much was learned in the
intervals between their appearances, so a lot of new material is
included in each subsequent book, and a lot of the older material is
not repeated.

This school pays homage to Sidney Coleman, a great scholar and
teacher, whose contributions to theoretical physics in general, and
the Erice schools in particular, are legendary. I always enjoyed
interacting with him. Although I did not know him very well, we did
cross paths in Aspen and elsewhere on a number of occasions. I
recall him once saying that there are three things that he does not
like, all of which are becoming popular: supersymmetry, strings, and
extra dimensions. Obviously, my views are quite different, but this
did not lessen my regard for him, nor did it harm our personal
relationship. In fact, I respected his honesty, especially as he did
not try to impose his prejudices on the community.

\section{Lecture 1: Overview of Superstring and M-Theory}

String theory arose in the late 1960s as a radical alternative to
conventional quantum field theory in which the fundamental objects
are one-dimensional strings rather than zero-dimensional points.
This proposal arose in the context of S-matrix theory, a subject
that has been much maligned, but whose imprint is indelible. From a
modern perspective, it is clear that theories of point particles and
theories of strings are related by dualities. Therefore the two
classes of theories, which used to seem completely at odds, actually
are deeply intertwined and do not have a sharp separation. Thus, I
would claim that the subject area that is currently called string
theory should be viewed as the logical completion of quantum field
theory. That being the case, string theory has a certain
inevitability, and there is nothing radical about it at all.

\subsection{Basic Concepts of String Dynamics}

Let us start by sketching what it means to construct a theory based
on strings. To do this it is convenient to review first
point-particle theory (\ie quantum field theory) from a
first-quantization viewpoint. This is not the way that it is usually
taught. The reason for doing this is that the analogous formulation
of string theory is much better understood, and much easier to
describe, than the string theory analog of second-quantized quantum
field theory.

For a {\it point particle} the classical motion makes the invariant
length of the world-line extremal. The corresponding action,
proportional to this length, is given by
$$
S = -m \int ds.
$$
Here $ds$ is the invariant line element, given in a general curved
spacetime by $ds^2 = g_{\m\n} dx^\m dx^\n$, and $m$ is the mass of
the particle in question. (We always set $\hbar =c =1$.) The motion
of a particle is given by a {\it world line} $x^\m (\t)$, where $\t$
is an arbitrary parameter for the trajectory. The action is
independent of the choice of this parametrization.

To pass to the quantum theory one computes an amplitude for
propagation from an initial spacetime point $x^\m_{\rm i} = x^\m
(\t_1)$ to a final spacetime point $x^\m_{\rm f} = x^\m (\t_2)$. As
Feynman taught us, this is given by the path integral (or sum over
histories)
$$
A_{\rm if} = \int_1^2 Dx^\mu(\t) \, e^{iS}.
$$
This is slightly oversimplified in one respect. The $\t$
reparametrization invariance is a type of gauge invariance that need
to be accounted for. One possibility, which is not always the most
convenient one, is to choose a gauge in which $\t$ is time, \ie
$x^0(\t) = \t$.

Interactions can be incorporated in this formalism by allowing world
lines to join or bifurcate and associating a coupling constant $g$
whenever this occurs. In this way, one can reproduce the
perturbation expansion of a second-quantized quantum field theory.
One shortcoming of this first-quantized approach is that it is not
very convenient for studying nonperturbative phenomena.

We can now ``invent'' string theory by doing the same thing for
one-dimensional extended objects.  If the string's topology is that
of a circle it is called a {\it closed string}. If the topology is
that of a line segment, it is called an {\it open string}. A string
sweeps out a two-dimensional surface in spacetime, called the {\it
world sheet} of the string. In the case of a closed string, the
topology of the world sheet is that of a cylinder, and in the case
of an open string it is that of a strip. For a {\it string} the
motion makes the invariant area of the world-sheet extremal
$$
S = -T \int dA .
$$
The invariant area element $dA$ is given by a simple formula
analogous to that of $ds$ in the point-particle case. The
coefficient $T$ is identified as the {\it tension} (or energy per
unit length) of the string. The world-sheet is described by
embedding functions $x^\m (\s, \t)$. Here $\s$ is interpreted as a
coordinate along the string (periodic in the case of a closed
string) and $\t$ is a timelike parameter. The choice of coordinates
is arbitrary in that there is now a two-dimensional
reparametrization invariance.

The basic idea of the extension to the quantum theory is the same as
before. Namely, one defines amplitudes for string propagation by a
path integral, properly taking account of the reparametrization
invariance. However, there are some very interesting differences
between the two cases: First of all, after covariant gauge fixing,
the world-sheet theory has a conformal symmetry. (The conformal
group in two dimensions is infinite dimensional.) Consistency of the
quantum theory requires cancellation of a conformal anomaly. The
most straightforward way to achieve this is to take the spacetime
dimension to be 26. This defines {\it critical string theory}.
Alternative approaches give {\it noncritical string theory}.

The second important difference from the point-particle case
concerns interactions. In the case of strings, these are uniquely
determined by the free string theory. They arise for purely a
topological reason. For example, the process in which a single
closed string turns into two closed strings is given by the {\it
pants diagram}. This is a smooth world sheet without any singularity
associated with the string junction. Such junctions need to be
incorporated in string path integrals. The smoothness of the world
sheet has a remarkable consequence: loop amplitudes have no
ultraviolet divergences! In the case of point-particle theories
these can be traced to the world line junctions, which are
short-distance singularities in the one-dimensional network of world
lines. Moreover, at least in the case of oriented closed strings,
there is just one Feynman diagram at each order of the perturbation
expansion!

The actions $S$ described above can be generalized to objects with
$p$ dimensions, called $p$-branes. However, the quantum analysis
breaks down for $p > 1$, because their world-volume theories are
nonrenormalizable. Later we will argue that various $p$-branes do
arise in string theories and M-theory as nonperturbative
excitations. The significance of this nonrenormalizability does not
concern the existence of $p$-branes. Rather, it means that they
cannot be treated as the fundamental objects on which to base a
perturbation expansion. The fact that this is possible for strings
is the feature that distinguishes them from higher-dimensional
$p$-branes. However, at strong coupling, when a perturbation
expansion is not helpful, this distinction evaporates and all branes
are more or less equal.

\subsection{A Brief History of String Theory}

\subsubsection{Hadronic String Theory}

String theory arose in the late 1960s in an attempt to understand
the strong nuclear force. This is the force that holds quark and
gluons together inside the strongly interacting particles (hadrons).
The story has a long and complex history that has not been assembled
yet in a single scholarly study. However, many of the participants
in this history presented their own personal recollections in a 2007
meeting at the Galileo Institute in Florence. My contribution to the
proceedings is available \cite{Schwarz:2007yc}.

Veneziano's discovery of a simple formula for a 2-particle to
2-particle scattering amplitude with many attractive features (such
as Regge behavior) quickly led to the realization that a theory
based on strings, rather than point-like particles, could account
for various observed features of the strong nuclear force
\cite{Veneziano:1968}. The original string theory, sketched in the
previous subsection, contains only bosons, and (as discussed above)
it is consistent for 26 spacetime dimensions. 25 of the dimensions
are spatial and 1 is time.

Another string theory containing fermions (as well as bosons) was
constructed  in 1971 by Pierre Ramond, Andr\'e Neveu, and me
\cite{Ramond:1971b,Neveu:1971b}. It requires 10 dimensions. Its
development led to the discovery of {\it supersymmetry}, a symmetry
that relates bosons and fermions. Strings in theories with this
symmetry are called {\it superstrings}.

The unrealistic dimension of spacetime was obviously a problem, but
not the only one. Another one was that the spectrum of string
excitations in both of the string theories includes {\it massless
particles}, whereas all hadrons have positive mass. In the early
1970s a better theory of the strong nuclear force, called {\it
quantum chromodynamics} (or QCD), was developed. As a result, string
theory fell out of favor. What had been an active community with
several hundred participants rapidly shrunk to a backwater of
theoretical physics kept alive by only a few diehards.

\subsubsection{Unification}

One of the massless particles in the closed-string spectrum of the
26-dimensional bosonic string theory, as well as in the
closed-string spectrum of ten-dimensional superstring theories, has
precisely the right properties to be the {\it graviton} -- the
particle responsible for transmitting the gravitational force. In
other words, it has spin two, which means that its polarization
tensor belongs to a symmetric and traceless representation of the
little group. Moreover, the requisite local symmetries are
incorporated, namely general coordinate invariance and local Lorentz
invariance. These properties ensure precise agreement with general
relativity at energies low compared to the string scale.

To be completely honest, it should also be mentioned that there is
also a massless scalar (called the dilaton). This field is very important,
because its vacuum expectation value determines the string coupling
constant. Somehow, the dynamics should give mass to this field to
avoid contradictions with experiments that show that the long-range
gravitational force is pure spin two to high accuracy. Presumably, this is
achieved at the same time as the extra dimensions are compactified. I will
discuss this more later.

The open-string spectrum contains massless spin one gauge fields, like those in
the standard model. Again, the usual gauge symmetries are incorporated,
which ensures agreement with Yang--Mills gauge theory at energies low
compared to the string scale.

For several years, string theorists (including myself) tried to
modify string theory so as to give mass to these fields, since we
knew that there are no massless hadrons. However, these attempts
always led to inconsistencies. Eventually it became clear that these
were essential features of the theory, and that we should let the
theory guide us, rather than trying to impose our will on it.
Accordingly, in 1974 Jo\"el Scherk and I proposed to use string
theory as a unified theory of all forces (including gravity), rather
than just the strong nuclear force \cite{Scherk:1974a}. Thus, we
stumbled upon an approach to achieving {\it Einstein's dream} -- a
unified theory of all fundamental forces. The connection to general
relativity at low energies was described independently by Yoneya
\cite{Yoneya:1973,Yoneya:1974}, though he did not propose using
string theory as a unified theory.

The interpretation of string theories as unified theories of gravity
and other forces has several advantages, which were immediately
apparent. All prior attempts to describe quantum corrections to
Einstein's theory of gravity assumed point particles. They gave
nonrenormalizable ultraviolet divergences, which makes the theory
unacceptable, at least within the context of perturbation theory.
String theory, on the other hand, was known to be UV finite to all
orders in perturbation theory. The intuitive reason for this, is
that the string world sheet is smooth, without any short-distance
singularities, no matter how complicated the topology.

A second advantage of using string theory as a theory of gravity is
that the extra dimensions can be an advantage rather than a disadvantage.
Extra spatial dimensions can be compact in a gravitational
theory, where the spacetime geometry is determined by the dynamics.
So the geometry can appear four-dimensional at low energies. It then
becomes a dynamical question whether the theory has acceptable quantum
vacua of this type.

\subsubsection{The Size of Strings}

When strings were supposed to describe hadrons their typical size
needed to be roughly that of a typical hadron, namely
$10^{-13} {\rm cm} $. However, in a relativistic
quantum theory of gravity there is a characteristic length, known as
the {\it Planck length}
$$
L_{\rm P} \sim \sqrt{\hbar G/c^3} \sim 10^{-33} {\rm cm}.
$$
This corresponds to a mass scale of about $10^{19}$ GeV.
This is the natural first guess for the fundamental string length scale
in a string theory containing gravity. This shrinkage by 20 orders of
magnitude was a very large conceptual change. Fortunately, the mathematics
remained pretty much the same. More recently, it has been appreciated that this
first guess might be modified for various reasons. There are possible corrections
involving factors of the coupling constant, the volume of compactified
extra dimensions, or even warpage in the geometry of extra dimensions.
Therefore, it makes sense for experimentalists to look for signs of the
string scale at the LHC, though I would be astonished if it did show up.

\subsubsection{The First Superstring Revolution}

String theory became a hot subject in the mid-1980s following some
important discoveries: anomaly cancellation \cite{Green:1984sg},
heterotic strings \cite{Gross:1984dd},
and Calabi--Yau compactification \cite{Candelas:1985en}.
By the time the dust settled, it appeared that there are exactly
five consistent superstring theories, each
of which requires {\it ten spacetime dimensions and
supersymmetry}. Three of them, called {\it Type I, Type IIA, Type IIB},
were introduced by Green and me \cite{Green:1981yb} building
on earlier work of Gliozzi, Scherk, and Olive \cite{Gliozzi:1977}.
The other two, called {\it $SO(32)$ heterotic and $E_8 \times E_8$ heterotic},
were formulated in \cite{Gross:1984dd}. They incorporate the gauge groups
identified in \cite{Green:1984sg}. The anomaly-free type I superstring
also has $SO(32)$ gauge symmetry.

Each of these five theories has no adjustable dimensionless
parameters. All dimensionless parameters arise either
\begin{itemize}
\item {\it dynamically} as the expectation values of scalar fields

or

\item as {\it integers that count something} such as topological
invariants, physical objects (branes), or quantized fluxes.
\end{itemize}
It appeared that the quest for a unique theory was almost at hand.
The only mystery was why there should be five theories when only one
is required.

One scheme looked particularly promising. Specifically, the $E_8
\times E_8$ heterotic theory has consistent vacuum solutions in
which six spatial dimensions form a compact {\it Calabi--Yau
manifold}. Calabi--Yau manifolds are a class of six-dimensional
manifolds that can be described as having three complex dimensions.
The precise mathematical statement is that they are K\"ahler
manifolds with $SU(3)$ {\it holonomy}.\footnote{ K\"ahler manifolds
are complex manifolds that have no torsion. The restriction to the
$SU(3)$ subgroup of the generic $U(3)$ holonomy group corresponds to
the vanishing of the first Chern class. It also implies the
existence of a covariantly constant spinor, which in turn implies
that some supersymmetry is preserved.}

The other four dimensions form Minkowski spacetime,
$M_{3,1}$. Thus, the ten-dimensional spacetime is a product space
$$
{\cal M}_{10} = {\rm CY}_6 \times M_{3,1}.
$$
Intuitively, one can imagine a Calabi--Yau space attached to every
point in four-dimensional spacetime. If the CY space is small, then
its structure cannot be probed at low energy. Experimental detection
requires energies greater than the inverse size of the compact CY
space. I will say more about these constructions in my second
lecture.

\subsubsection{The Second Superstring Revolution}

In the mid-1990s there was a remarkable burst of progress in addressing
some of the issues that we have raised. A few of the key initial contributions
were \cite{Sen:1994fa,Hull:1994ys,Witten:1995ex}. The main lessons were
the following:
\begin{itemize}
\item There is just one theory! What had been viewed as
five theories are actually five different corners of a space of
solutions to a unique underlying theory. The five superstring
theories are related by various surprising equivalences, called
dualities:

{\it T-duality} ($R \to 1/R$) relates the two type II superstring
theories, and also the two heterotic string theories. Thus, for
example, if the type IIA superstring theory is considered in a
spacetime that is nine-dimensional Minkowski spacetime times a
circle of radius $R$ ($M_{8,1} \times S^1$), this is equivalent to
the type IIB superstring theory in a spacetime that is
nine-dimensional Minkowski spacetime times a circle of radius $1/R$.
(Here we are using units in which the fundamental string length
scale $L_{\rm s}$ has been set equal to one.) Thus, the two type II
theories in ten-dimensional Minkowski spacetime are two limiting
cases of a continuum of consistent possibilities that are connected
by letting the radius $R$ range from zero to infinity. This radius
is determined by the vacuum expectation value of a modulus field
(one of the components of the ten-dimensional metric).

{\it S-duality} ($g_{\rm s} \to 1/g_{\rm s}$) relates the type I
superstring theory and the $SO(32)$ heterotic string theory. The
string coupling constant $g_{\rm s}$ is given by the vacuum
expectation value of a modulus field called the {\it dilaton}. Prior
to this discovery one only knew how to analyze string theory at weak
coupling using perturbation theory. S-duality implies that this pair
of theories is continuously connected by varying $g_{\rm s}$ from
zero to infinity. It also implies that the strong coupling expansion
of one theory is given by the weak coupling expansion of the other
theory. The type IIB superstring theory is related to itself in a
similar manner, which determines its strong coupling expansion as
well.

\item Knowing the strong coupling expansion of three of the five
superstring theories, it was natural to explore the strong coupling
behavior of the other two superstring theories. The conclusion is
quite striking. At strong coupling the type IIA and the $E_8 \times
E_8$ theories become 11-dimensional. More precisely, the eleventh
dimension has size proportional to $g_{\rm s}L_{\rm s}$. In the IIA
case it is circle whereas in the $E_8 \times E_8$ case it is a line
interval with two endpoints. In the latter case the 11-dimensional
spacetime has two ten-dimensional faces. It turns out that one $E_8$
gauge group is localized on one face and the other $E_8$ on the
other face, so that they are spatially separated.

In either the type IIA or $E_8 \times E_8$ theory the $g_{\rm s} \to
\infty$ limit leads to the same 11-dimensional Minkowski spacetime
theory. This 11-dimensional theory, which is still not yet very well
understood, is called {\it M-theory}. M-theory is approximated at
low-energies by 11-dimensional supergravity, a theory that was
formulated many years earlier \cite{Cremmer:1978km}.

\item New stable objects, called {\it $p$-branes}, arise
nonperturbatively. $p$ is the number of spatial dimensions they
occupy. Thus, in this notation, a point-particle is a 0-brane, a
string is a 1-brane, and so forth. $p$-branes can carry generalized
conserved charges that are sources for antisymmetric tensor gauge
fields with $p+1$ indices, which generalize the Maxwell field in the
$p=0$ case.\footnote{These gauge fields can also be regarded as
differential forms $A = \frac{1}{(p+1)!}A_{\m_1 \m_2 \cdots
\m_{p+1}} dx^{\m_1} \wedge dx^{\m_2}\wedge \ldots \wedge
dx^{\m_{p+1}}$. In this notation the coupling of the gauge field to
the brane world volume is given by the simple expression $\m \int
A$. The parameter $\m$ is the brane charge.} The conserved charges
satisfy generalized Dirac quantization conditions. In $D$ spacetime
dimensions, the magnetic dual of a $p$-brane is a $(D-p-4)$-brane.

The main categories of $p$-branes are called D-branes, M-branes, and
NS-branes. D-branes are characterized by Dirichlet boundary
conditions for open strings. In the type IIA theory stable D-branes
exist for even values of $p$ and in the type IIB theory they exist
for odd values of $p$. M-theory has a three-form gauge field, which
can couple electrically to an M2-brane and magnetically to an
M5-brane. The NS5-brane is the magnetic dual of the fundamental
string in the heterotic and type II superstring
theories.\footnote{The type I superstring does not carry a conserved
charge, and therefore it can break. It does not have a magnetic
dual.}

\end{itemize}

\subsection{AdS/CFT: Holographic Duality}

In 1997 Maldacena proposed a new class of dualities relating certain
string theory and M-theory solutions in anti de Sitter
geometries\footnote{Anti de Sitter space is a maximally symmetric
spacetime with constant negative curvature.} (AdS${}_{d+1}$ times a
compact space) to conformally invariant quantum field theories
(CFT${}_{d}$) \cite{Maldacena:1997re}. The first piece of evidence
for such a relationship is the fact that the conformal symmetry
group in $d$ dimensions, $SO(d,2)$, is the same as the isometry
group of anti de Sitter space in $d+1$ dimensions. The precise
relationship between amplitudes/correlation functions in the two
pictures was spelled out \cite{Gubser:1998bc,Witten:1998qj}. Such
dualities are called {\it holographic}, because string theory in
AdS${}_{d+1}$, which has $d+1$ dimensions, is equivalent (dual) to a
conformally invariant quantum field theory in $d$ dimensions: the
dual CFT${}_{d}$. This is reminiscent of recording a
three-dimensional image on a two-dimensional emulsion. The reason
that two constructions that look so radically different can
nonetheless be equivalent is that when one of them is weakly
coupled, the other is strongly coupled. If it were easy to
understand both of them at the same time, there would be a paradox.

This is an enormous subject, and I will just say a little bit about
it here. Three maximally symmetric examples are given in Maldacena's
original paper: Type IIB superstring theory in an $AdS_5 \times S^5$
background geometry with $N$ units of five-form flux threading the
sphere is dual to ${\cal N}=4$ super Yang--Mills theory with an
$SU(N)$ gauge group. This by far the most studied example. The
leading term in the 't Hooft large-$N$ expansion of the gauge theory
(the planar approximation) is supposed to be dual to the string
theory in the tree approximation. This much of the duality is quite
well understood, and I think it is likely that it will be completely
proved someday. The higher-order terms in the $1/N$ expansion
correspond to string loop corrections. This duality can be
generalized to examples with less supersymmetry in which the
five-sphere is replaced by another five-dimensional {\it
Sasaki--Einstein space}. It is also possible to replace the $AdS$
space by a space that is only asymptotically anti de Sitter. In this
case the conformal field theory is perturbed by relevant operators.

The other two maximally supersymmetric examples relate M-theory in
an $AdS_4 \times S^7$ background geometry with flux to a
three-dimensional conformal field theory and M-theory in an $AdS_7
\times S^4$ background geometry with flux to a six-dimensional
conformal field theory. Within the past year there has been a great
deal of progress in understanding the first case (as well as an
orbifold generalization) \cite{Aharony:2008ug}. On the other hand,
there has been very little progress in understanding the second
M-theory duality.

The general AdS/CFT framework is being used to construct brane
configurations that capture many of the essential features of QCD,
even though it isn't precisely the same theory \cite{Sakai:2004cn}.
From the modern viewpoint, such constructions can be regarded as
spinoffs of string theory rather than its central goal. Even so,
this brings string theory back to its historical origins. It also
makes the reasons for the early failure to achieve the original goal
of describing hadrons much clearer. For one thing, it is now
apparent that a string theory dual of QCD must involve at least one
extra spatial dimension \cite{Polyakov:1998ju}. There are also
spinoffs to other fields: 1) There have been striking applications
of AdS${}_5$/CFT${}_4$ duality to studies of the quark-gluon plasma
(RHIC, LHC), especially the computation of the viscosity to entropy
density ratio $\eta/s$ \cite{Policastro:2001yc,Kovtun:2004de}.
Taking the field theory at finite temperature corresponds to
including a black hole in the dual AdS geometry! 2) Similar
constructions could have condensed matter applications. Systems that
show promise are ones in which there is a phase transition at which
the conformal field theory is strongly coupled. Then the dual
black-hole description could be helpful. A possible application of
this type is to high $T_c$ superconductors. For a recent overview
see \cite{Sachdev:2008ba}. 3) Nonrelativistic versions of AdS/CFT
may be relevant to Bose-condensed cold-atom systems under special
conditions \cite{Balasubramanian:2008dm,Son:2008ye}.

\subsection{Some Other Topics in String Theory}

\subsubsection{Brane Worlds}

As we already mentioned, the defining property of D-branes is that
strings can end on them. Away from D-branes, a string must have the
topology of a circle, but the string can break provided that its
ends are attached to D-branes. The lowest excitation mode of the
open string is a massless gauge field. This has the crucial
consequence that Yang--Mills gauge theories (like the standard
model) can live on stacks of D-branes. For example, in the case of
$N$ coincident type II D-branes, the world-volume theory of the
branes is a supersymmetric $U(N)$ gauge theory.

This picture raises new possibilities for the physics of the extra
dimensions, which is rather different from the usual Kaluza--Klein
picture. Rather than having a wave function that is uniformly spread
over the compact extra dimensions, a low-energy field can have its
wave function concentrated on (or close to) a D-brane, which is a
defect in the extra dimensions that fills the noncompact spacetime
dimensions. The simplest such possibility is that the observable
Universe is actually a stack of D3-branes, which are points embedded
in the six extra spatial dimensions. A generalization of this
approach uses intersecting stacks of higher-dimensional D-branes.
For example, one could consider stacks of D6-branes wrapping
3-cycles in the six extra dimensions.

\subsubsection{Flux Compactifications}

As has already been mentioned, the type II superstring theories
contain various massless antisymmetric-tensor (or differential-form)
gauge fields
$$
A_n = \frac{1}{n!} A_{\m_1 \m_2 \ldots \m_n} dx^{\m_1} \wedge
dx^{\m_2} \wedge \ldots \wedge dx^{\m_n}.
$$
$p$-branes with $p =n-1$ are sources for these fields.
These differential-form gauge fields have gauge-invariant field
strengths of the form $F_{n+1} = dA_n,$ which are invariant under
gauge transformations of the form $\d A_n = d \L_{n-1}$. This is a
straightforward generalization of Maxwell theory (the $n=1$ case).

If the compact dimensions contain nontrivial $(n+1)$-cycles,
$C_{n+1}$, it is possible to have (quantized) flux threading the
cycle:
$$
\int_{C_{n+1}} F_{n+1} = 2\pi N, \quad \quad N\in \IZ.
$$
Such flux can be present without a $p$-brane source. The possibility
of constructing superstring vacua containing such fluxes greatly
increases the number of possible vacua, even though there are
constraints that must be satisfied. Mike Douglas analyzed one
particular CY compactification of the type IIB theory and estimated
the number of distinct flux vacua to be about $10^{500}$
\cite{Douglas:2003um}. For a review of this subject, see
\cite{Douglas:2006es}.

The {\it moduli problem} --- the occurrence of massless scalar
fields $\phi_i$ with continuously adjustable vacuum expectation
values --- can be solved in the context of flux compactification:
The fluxes induce a nontrivial potential energy function for the
moduli $V(\phi_i)$. This {\it landscape} function has isolated
minima, which are what Douglas counted. The moduli are massive at an
isolated minimum.

\subsubsection{Warped Compactification}

One of the important properties of flux compactifications is that
they give rise to warped geometries. This means that the
ten-dimensional geometry is no longer a direct product of the form
$${\cal M}_{10} = M_{3,1} \times K_6.$$
Instead, the four-dimensional Minkowski part of the ten-dimensional
metric is multiplied by a {\it warp factor} $h(y)$ that depends on
the position $y$ in the internal manifold
$$
ds^2_{10} = h(y) dx \cdot dx + ds^2_6.
$$

In 1999 Randall and Sundrum proposed that an exponential $y$
dependence could give a large ratio between the warp factors at the
positions of two 3-branes, which they called the {\it standard model
brane} and the {\it Planck brane} \cite{Randall:1999ee}. They
proposed that this could solve the hierarchy problem. This kind of
exponential warping is exactly what one has in the case of an AdS
spacetime. The new idea was to only consider a slice of AdS
spacetime between two branes. This is only an approximation to what
can be achieved in honest string theory solutions. Nonetheless, the
RS scenario can be made rather precise in the context of flux
compactifications with branes added \cite{Giddings:2001yu}. If this
is how Nature works, quantum gravity effects might be accessible at
the LHC. The experimentalists intend to look for them.

\subsection{String Cosmology}

An important question, which was not discussed much in the 20th century, is
``What does string theory have to say about cosmology?'' Nowadays,
string cosmology is a very active research area. There are even
entire conferences devoted to it. Clearly, progress in this area
requires understanding {\it time-dependent} solutions of string theory,
which is technically challenging.

There is a lot of evidence that the very early Universe underwent a
period of inflation during which the scale factor grew exponentially
by a factor of at least $e^{60}$ (``60 e-foldings''). In simple field
theory models this is described in terms of a ``slowly rolling''
scalar field called an {\it inflaton}.
Proposals for the string-based origins of inflation, or possible
alternatives, are being explored extensively. One scenario is based
on CY compactification with flux and warped throats in the geometry
\cite{Kachru:2003aw,Kachru:2003sx}. In this scenario inflation takes
place as a D3-brane moves down a throat, attracted to an
anti-D3-brane at the bottom until they {\it collide and annihilate}.
A scalar mode of an open string connecting the branes is the
inflaton. The annihilation releases the brane tension energy. It
heats up the Universe to start the {\it hot big bang epoch}. All
sorts of strings are produced, and some might survive to be
observable as {\it cosmic superstrings} \cite{Polchinski:2007qc}.

\section{Lecture 2: Superstring Phenomenology}

Since the focus of this school is the physics of the LHC, it seems
appropriate to discuss attempts to relate string theory to the
particle physics phenomenology in greater detail than other topics.
That is the purpose of this lecture. However, before getting into
the details of such efforts, let me emphasize that the unique feature of
string theory is that it unifies gravitation and particle physics.
Moreover, its most successful prediction (or postdiction, if you wish)
so far is that {\it Gravity Exists!}

It is not yet clear how predictive string theory is for particle
physics. The discovery of the {\it string theory landscape} has led
to some pessimism in this regard. However, as I will argue, there
are good reasons to believe that if one makes certain experimentally
motivated assumptions and inputs some experimental details, string
theory could be quite predictive. I don't think it is reasonable to
expect to predict all experimental facts from first principles. It
would already be a great success, if one could predict a large
number of facts starting from a small number of experimentally
motivated assumptions. How far one can go in this direction remains
to be seen.

In perturbative string theory there are two basic ways to obtain
non-abelian gauge symmetry in four dimensions: 1) It is already
present in 10 dimensions (type I and heterotic strings) and broken
partially by compactification. 2) It arises as the symmetry of
compact extra dimensions (Kaluza--Klein). The Kaluza--Klein approach
does not give anything realistic. Nonperturbatively, there are other
possibilities that we will discuss later.

\subsection{Perturbative Heterotic String}

The gauge groups of the standard model and grand unified models
embed nicely in $E_8$ but not in $SO(32)$. In the period 1985--95
only one scheme looked promising for phenomenology: Calabi--Yau
compactification of the $E_8 \times E_8$ heterotic string. This
entails taking the ten-dimensional geometry to be a product of
four-dimensional Minkowski spacetime and a six-dimensional
Calabi--Yau space, $ {\cal M}_{10} = {\rm CY}_6 \times M_{3,1}$.
There must also be nontrivial gauge fields, satisfying certain
consistency conditions (related to anomaly cancellation),
in the internal space. This gives a GUT-like
effective four-dimensional theory at low energy with ${\cal N} = 1$
supersymmetry. These effective theories have a number of attractive
features:

\begin{itemize}
\item They have the structure of {\it supersymmetric grand unified theories}.
The advantages of low-energy supersymmetry and grand unification are
therefore naturally incorporated.

\item Each solution has a definite number of families
of quarks and leptons determined by the topology of the CY space.

\item The standard model gauge symmetry can be embedded in one $E_8$
factor. Then there is a {\it hidden sector}, associated to the
second $E_8$ factor. Supersymmetry can break dynamically in the
hidden sector. This breaking is communicated by singlet fields,
which have gravitational strength interactions, to the visible
sector. This is usually called {\it gravity mediation}, though more
than gravity is involved.

\item There are several good {\it dark-matter} candidates:
The lightest supersymmetric particle (or LSP) is absolutely stable
in schemes with unbroken R-parity. In gravity mediation the leading
dark-matter candidates are the lightest neutralino (a mixture of the
partners of the photon, $Z$ boson, and Higgs bosons) and
hidden-sector particles.\footnote{In other mediation schemes, such
as gauge mediation, the gravitino (the supersymmetry partner of the
graviton) and saxions (supersymmetry partners of axions) are
candidates.}
\end{itemize}

These developments created a lot of excitement in the mid-80s. In my
view a large part of the theory community made a phase transition
from being too skeptical of string theory to being too optimistic
about the short-term prospects for constructing a realistic theory.
The successes were qualitative, and there were many problems and
puzzling questions:

\begin{itemize}
\item Hundreds of Calabi--Yau manifolds were known (now many
thousands). Which one of them, if any, is the {\it right one}? Is
there a theoretical principle for deciding?

\item Why are there four other consistent superstring theories?
Can they give interesting solutions, too?

\item The CY compactification scenario described above is
based on perturbation theory. What new {\it nonperturbative
features} appear at strong coupling?

\item These CY solutions typically give a large number of massless
scalar fields (called {\it moduli}).\footnote{Moduli fields have a
flat potential. This means that the vacuum energy does not depend on
their values.} The moduli fields have gravitational strength
interactions, but long-range scalar forces of that strength are
ruled out by standard tests of general relativity. Thus, all these
CY solutions are unsatisfactory unless there is some way of giving
mass to the moduli. (Recall that in the first lecture I reported
that this can be achieved in flux compactifications of type II
superstring theories.)

\item In a gravitational theory the value of the minimum of the
potential also matters, because it determines the {\it vacuum energy
density}, which is the effective cosmological constant. For many
years it was assumed (without a good theoretical argument) that this
should be zero. However, about a decade ago it was discovered in
studies of distant supernovas (and confirmed by other means) that
the vacuum energy density (or dark energy) is very small and
positive. The value is roughly $10^{-120}$ in Planck units, which is
the smallest measured number in all of science. Typically, one would
expect to compute a result of order unity, so it is a great
challenge to try to account for this result.\footnote{If the true
value were zero, and there were no dark energy, this would still be
a big problem.} Supersymmetry can help, but once it is broken at the
TeV scale, one still expects to obtain a value of order $10^{-60}$
in Planck units, which is 60 orders of magnitude too large.

\item The ratio of the Planck scale to the GUT scale comes out an
order of magnitude too small.

\item Extra family-antifamily pairs typically occur.

\end{itemize}

\subsection{Other Approaches}

Following the second superstring revolution in the mid-1990s several
new approaches to particle phenomenology opened up. These exploit
discoveries such as

\begin{itemize}
\item 11d M-theory

\item Singular geometries (e.g. orbifolds and orientifolds)

\item D-branes and other branes

\item Flux compactifications and warped geometries

\item Nonperturbative type IIB superstring solutions (F-theory)

\end{itemize}

\subsubsection{Heterotic M-Theory}

Ho\v{r}ava and Witten discovered that the $E_8 \times E_8$ heterotic
string at strong coupling is 11-dimensional \cite{Horava:1995qa}. In
particular they showed that anomaly cancellation requires that $E_8$
gauge fields live on each boundary component \cite{Horava:1996ma}.
In heterotic M-theory the geometry is a solid slab with two
ten-dimensional boundaries. One $E_8$ lives on each boundary.

Calling the coordinate that runs across the slab $y$, the shape and
size of the $CY_6$ manifold can be $y$-dependent. In this way one incorporates
nonperturbative improvements of the perturbative heterotic constructions.
Taking this possibility into account improves the range of
possible ratios of the Planck scale to the GUT scale \cite{Witten:1996mz}.
Fairly realistic examples have been constructed by the Penn group.
(See \cite{Bouchard:2005ag,Braun:2005nv} and references
therein). However, despite their many realistic features, these examples
still have massless moduli.

\subsubsection{M-Theory on a $G_2$ Manifold}

The strong coupling limit of type IIA superstring theory is eleven-dimensional
M-theory, which is approximated at low energies by 11-dimensional supergravity.
To obtain ${\cal N}=1$ supersymmetry in four dimensions at low energies
requires a compactification of the form
$$
{\cal M}_{11} = K_7 \times M_{3,1},
$$
where $K_7$ is a compact 7-manifold of $G_2$ holonomy. ($G_2$ is the
smallest of the five exceptional Lie groups. It has 14 generators.)
If the $G_2$ manifold is smooth, then the effective four-dimensional
theory at low energy has no non-abelian gauge symmetry or parity
violation. To overcome this problem it is necessary to consider
$G_2$ manifolds with certain types of singularities
\cite{Acharya:2001gy}. While the necessary types of singularities
are well understood, there has been little progress on constructing
compact 7-manifolds of $G_2$ holonomy that contain such
singularities. Such manifolds, having an odd dimension, are
obviously not complex manifolds. The math is much tougher when
techniques of complex analysis cannot be used. Thus no
quasi-realistic models of this type have been constructed.
Nonetheless, it is possible to make some general statements, as in
\cite{Acharya:2008hi}, for example.

\subsubsection{Intersecting D-branes in Type II}

Consider Type IIA superstring theory compactified on a six-torus.
Suppose that a stack of $M$ D6-branes wraps a 3-cycle of the torus
and fills the four-dimensional spacetime. This gives a
supersymmetric $U(M)$ gauge theory. Suppose now that a second stack
of $N$ D6-branes wraps a different 3-cycle that intersects the first
one at $n$ points. Then the four-dimensional theory has $U(M) \times
U(N)$ gauge symmetry, and it contains $n$ bifundamental chiral
matter multiplets transforming as $(M,\overline N)$.

It is striking that constructions that are this simple
can capture roughly the types of groups and representations that appear in the
standard model.  Moreover, here are generalizations involving orbifolds and
orientifolds that are quite realistic. But, even then, there are
some challenging problems that remain:
\begin{itemize}
\item Not easy to get the correct $U(1)$ of weak hypercharge and no
other $U(1)$ groups at low energy. Some $U(1)$s are lifted by a
four-dimensional version of the anomaly cancellation mechanism. However,
this does not automatically guarantee that just the
desired $U(1)$ group survives.

\item There are many massless moduli. Localizing the branes
at singularities eliminates some of them. However, it is hard
to remove all of them, as is required. One approach to overcoming
this problem is to have the brane intersections be localized at
singularities of the geometry. As an example of this approach
see \cite{Verlinde:2005jr}.

\item This type of scheme does not incorporate grand unification.
It is not certain that we should, since the hints of grand unification could
prove to be illusory. However, the more conservative guess is that
it is not an illusion.
\end{itemize}

For reviews of this remarkable subject see
\cite{Blumenhagen:2005mu,Blumenhagen:2006ci,Malyshev:2007zz}.

\subsubsection{Type II on a Calabi--Yau with Flux}

Calabi--Yau compactification of Type II superstring theories
$$
{\cal M}_{10} = {\rm CY}_6 \times M_{3,1}
$$
give ${\cal N}  =2$ supersymmetry in four dimensions. Also, it does not
give non-abelian gauge symmetry or chiral matter. So, without further
measures, this is completely unrealistic.
By including D-branes and/or flux one can overcome these
difficulties. In fact, as discussed in the first lecture,
the flux even helps to stabilize moduli. It
also gives warping of the geometry, as was also discussed.

This approach allows one to embed Randall--Sundrum models in a
string-theoretic setting. Recall that RS proposed that the warping
could used to solve the hierarchy problem and that it could
make quantum gravity effects accessible at LHC energies.
In this setting one can break supersymmetry by adding anti-D3-branes.
As we discussed, this approach leads to interesting string-theoretic
models of inflation in which the origin of the big bang is described as
a brane-antibrane collision. However, it has not yet yielded
quasi-realistic particle physics models as far as I am aware.

The very large number of vacua that typically arise in flux compactifications 
may have discouraged some people from trying to construct viable particle 
physics models. However, there is still a considerable effort underway
exploring the possibilities for constructing realistic examples.
See, for example, \cite{Lust:2004fi,Marchesano:2004xz}.
This large number of vacua has motivated the suggestion
that various parameters of Nature (such as the cosmological constant)
should be studied statistically on the landscape. This approach seems to
assume implicitly that Nature corresponds to a more or less random vacuum.
This in turn is motivated by some notion about how Universes are
spawned in the Multiverse in a process of eternal inflation. The
story gets even more entangled when the {\it anthropic principle}
is brought into the discussion. Some people are enthusiastic about this
type of reasoning, but I find it fundamentally defeatist. Even though 
in some sense it may
be correct, I remain optimistic that there is a more reductionist way of 
understanding particle physics. The next section describes an approach
that is more in that spirit.

\subsection{F-Theory Local Models}

\subsubsection{Local Versus Global Models}

In {\it global models} one fully specifies the compactification
geometry and all background fluxes and other fields. This completely
determines a vacuum and hence the physics at all scales -- up to and
beyond the Planck scale. The heterotic string constructions that we
have discussed are examples of global models.
{\it Local models} are less ambitious. They are supposed to
describe particle physics in a limit where gravity is turned off.
More precisely, the ratio of the Planck scale to the unification scale,
$M_{\rm Pl}/M_{\rm GUT}$ can be made arbitrarily large.

In certain classes of models in which matter fields are restricted
to branes, the low-energy physics only depends on the geometry in
the vicinity of the branes (in the extra dimensions). If one can
decompactify the extra dimensions in directions transverse to the
branes, this has the effect of increasing the four-dimensional
Planck mass while keeping other scales fixed. Then one only needs to
focus on the geometry in the immediate vicinity of the branes losing
sensitivity to features in the geometry that are far away from the
relevant branes. This approach is sometimes called {\it geometric
engineering}.

Such local models are more amenable to a ``bottom-up'' construction.
The basic idea is that one should first try to ``engineer'' a
phenomenologically viable local model that correctly describes all
nongravitational physics up to the unification scale to good
accuracy. Then one should try to find a ``global completion'' of
this local model that fully describes the geometry of the compact
dimensions and brings gravity back into the mix. It is plausible
(but not demonstrated, as far as I know) that the vast majority of
local models do not have a global completion. Thus the existence of
a global completion would be a confirmation that the original local
model was on the right track. Furthermore, the global completion is
likely to require that parameters (and other features) are
correlated even though they were not correlated in the original
construction of the local model. Altogether, one could imagine that
the construction of a viable local model would be somewhat
predictive, and that the construction of its global completion would
be more predictive.

\subsubsection{What is F-Theory?}

Type IIB superstring theory \cite{Green:1981yb} is the setting for
F-theory. Type IIB superstring theory is also the
framework where most of the recent progress in flux compactifications,
moduli stabilization, and SUSY breaking has been made.
F-theory, which utilizes nonperturbative type IIB vacua containing
7-branes, was introduced in 1996 by Vafa \cite{Vafa:1996xn}.
F-theory configurations that give four-dimensional models
contain various 7-branes wrapping compact 4-cycles of the
six-dimensional compact space $B$. They also fill the four-dimensional
Minkowski spacetime.

The type IIB theory in ten-dimensional Minkowski spacetime
contains two massless scalar fields, often called $C_0$ and
$\phi$, which are conveniently combined to form a complex scalar field
$$
\t = C_0 + i e^{-\phi}.
$$
The field $C_0$ is a Ramond-Ramond zero-form, which couples
electrically to D-instantons and magnetically to D7-branes. It
behaves, at least in some respects, like a ten-dimensional analog of
an axion. The field $\phi$ is called the {\it dilaton}. It has the
special property that the value of $e^{\phi}$ is the string coupling
strength, \ie $g_{\rm s} = \langle e^\phi \rangle$.

Type IIB superstring theory has a {\it nonperturbative
discrete gauge symmetry} given by \cite{Hull:1994ys}
$$
\t \to \t' = \frac{a\t + b}{c\t + d}, \quad {\rm where}
\quad \left(\begin{array}{cc}a&b\\c&d\\ \end{array}\right) \in SL
(2, {\IZ}).
$$
This means that $a,b,c,d$ are integers satisfying $ad-bc=1$. Other
fields transform at the same time. A special case is the
transformation $\t \to - 1/\t$. For $C_0 = 0$ this maps $\phi \to
-\phi$ and hence $g_{\rm s} \to 1/g_{\rm s}$, which is S-duality.
The group $SL(2, \IZ)$ is also the modular group of the torus. As
often happens in string theory, things that look the same actually
are the same. In this context this means that $\t$ can be
interpreted as the complex structure of a torus. So the picture is
that to every point $y \in B$ one associates a torus with the
complex structure $\t(y)$. As $\t$ approaches certain special values
the torus becomes singular. (Mathematicians say that the torus {\it
degenerates}.) This happens generically on the 4-cycles where the
7-branes are located.

When one analytically continues $\t(z)$ along a curve encircling a
7-brane it comes back to a transformed value given by an $SL(2,\IZ)$
transformation. The 7-branes are classified by these transformations
(more precisely the conjugacy classes), so there are an infinite
number of distinct types of 7-branes. They can be described by a
pair of integers $(p,q)$ without any common divisors. Type IIB
superstring theory also has an infinite number of different types of
strings with a similar classification scheme. In fact, a $(p,q)$
7-brane is one on which a $(p,q)$ string can end. A special case of
this statement is that a fundamental string can end on a D7-brane.
The associated $SL(2,\IZ)$ transformation in this case is $\t \to \t
+1$. In this case the dilaton is constant and only $C_0$ varies, so
it is possible to have weak coupling and use perturbative string
techniques.

For a consistent F-theory solution, the varying field $\t(z)$ and
the compact six-dimensional space $B$ are required to define an {\it
elliptically fibred Calabi-Yau four-fold} (with section). Then, the
low-energy effective four-dimensional theory has ${\cal N} =1$
supersymmetry. One subtlety is that this construction makes it seem
as if the compact space is eight-dimensional, since that is the
dimension of a Calabi-Yau four-fold. If this were the case the
entire spacetime dimension would be 12. However, this is definitely
not the case. The tori that describe the fibres have a complex
structure, given by $\t(z)$, but they have zero area. This can be
understood as a limit of tori with finite area, but a complete
explanation requires invoking a duality relating F-theory to
M-theory, which I will not present here. There had been a
considerable amount of progress in understanding F-theory before
this year \cite{Friedman:1997yq,Bershadsky:1997zs,Andreas:1999ng},
but it was not clear how to go about making realistic models.

\subsubsection{Construction of Local F-Theory Models}

The F-theory approach to phenomenology has been pursued this year by
various groups starting with Donagi and Wijnholt and Beasley, Heckman, 
and Vafa \cite{Donagi:2008ca} -- \cite{Heckman:2008qa}.  
The new proposal, which has given the
subject a new lease on life, is to focus on models in which one can
define a limit in which gravity is decoupled. The criterion is that
it should be possible to make the dimensions transverse to the
4-cycles wrapped by the 7-branes arbitrarily large. Equivalently, it
should be possible to contract the 4-cycles to points while holding
the six-dimensional volume fixed. Such contractible 4-cycles must be
positive curvature K\"ahler manifolds. These are fully classified
and are given by manifolds called {\it del Pezzo manifolds} (or del
Pezzo surfaces), which are denoted dP${}_n$. The integer $n$ takes
the values $0\leq n \leq 8$.\footnote{There is also one other
possibility, which is a product of two two-spheres.} The del Pezzos
have a close relationship with the exceptional Lie algebras $E_n$.
The basic idea is that they contain 2-cycles whose intersections are
characterized by the $E_n$ Dynkin diagram. By this type of F-theory
construction, one can construct an $SU(5)$ or $SO(10)$ SUSY-GUT
model.

Constructions that involve 7-branes of various types are much more
subtle -- and also more interesting than ones that only involve
D7-branes. D7-branes are mutually local. A stack of $N$ of them
gives $U(N)$ gauge symmetry. Matter fields at intersections (due to
stretched open strings) are bifundamental. However, different kinds
of 7-branes are mutually nonlocal. As a result, there are stacks
(corresponding to the ADE classification of singularities) that can
give $U(N)$, $SO(2N)$ or even $E_N$ gauge symmetry. Exceptional
gauge groups, spinor representations of $SO(10)$, and nonzero
top-quark Yukawa couplings in $SU(5)$ are possible for F-theory
vacua. In constructions that only involve D7-branes these phenomena
can only arise nonperturbatively. Since the top quark Yukawa
coupling must be quite large to account for the large top quark
mass, the F-theory viewpoint seems preferable.

The crucial assumption of the local F-theory approach -- the
existence of a decoupling limit in which four-dimensional gravity
can be turned off -- may or may not be physically correct. The fact
that $M_{\rm Pl}/M_{\rm GUT}$ is in the range 100--1000, depending
on the precise definition of each of these mass scales, makes it
plausible. There are an infinite variety of types of 4-cycles that
7-branes could wrap if one did not make this assumption. Therefore
this principle picks out a tiny corner of the string theory
landscape. In this way, the doom and gloom that the discovery of
this landscape has generated is swept aside, at least if the basic
assumption is correct.

In particular, it allows one to construct an $SU(5)$ or $SO(10)$
SUSY-GUT with three families. The $SU(5)$ case is more
straightforward. There are three basic rules:

\begin{itemize}
\item Each stack of 7-branes on a 4-cycle $S$ determines a gauge
group. One stack should give the GUT gauge group. Other ``matter
7-branes'' are also needed.

\item Pairs of 4-cycles intersect on 2-cycles $S\cap S'$. Each such
2-cycle is a Riemann surface that has associated {\it
chiral matter fields} given by zero-modes of the Dirac operator.
The number of such zero modes determines the number of families.

\item Three 2-cycles can intersect at points. Products of wave functions
at these intersections determine the {\it Yukawa couplings}.

\end{itemize}

These three rules can be summarized by saying that the gauge fields
live in eight dimensions, the matter fields live in six dimensions,
and the interactions take place in four dimensions. Gravity, which
is ignored in a local construction, lives in ten dimensions.

\subsubsection{Gauge Symmetry Breaking}

One of the strengths of the local F-theory approach to model
building is the way that symmetry breaking can be incorporated. In
conventional SUSY-GUTS based on four-dimensional quantum field
theory, the breaking of the $SU(5)$ GUT symmetry to the $SU(3)
\times SU(2) \times U(1)$ standard model gauge group requires the
introduction of Higgs fields belonging to large representations of
$SU(5)$. This is definitely ugly, and does not work very well. In
string theory such representations are not present in the low mass
spectrum, but there are other available mechanisms. In the case of
the heterotic string constructions the standard method introduced in
\cite{Candelas:1985en} is to associate nontrivial Wilson lines to
noncontractible cycles of the Calabi--Yau geometry.

In the local F-theory
models there are no noncontractible cycles and a different mechanism
for breaking the $SU(5)$ gauge symmetry to the standard model gauge
group is required. Fortunately, one is available. The proposal of Vafa
and collaborators is introduce nonzero flux. In doing this one has
to choose the $U(1)$ subgroup of $SU(5)$ that this flux corresponds to.
By choosing it to correspond to the weak hypercharge $U(1)$ subgroup
one gets exactly the desired symmetry breaking pattern. This flux
is called {\it hyperflux} in the recent literature. One also has to
choose the 2-cycle inside the del Pezzo surface that carries this flux.
This is an elaborate and beautiful story that I don't intend to explain
in detail here. One crucial point is that by making appropriate
choices it is possible to arrange to obtain complete $SU(5)$ multiplets
when one wants them and incomplete multiplets when that is desired.
In the case of the quarks and leptons one wants complete ${\bf\bar 5}$
and {\bf 10} multiplets. In the case of the Higgs bosons one wants
only the doublet pieces of the {\bf 5} and ${\bf\bar 5}$ multiplets,
eliminating the triplets, which would give rapid proton decay. All this
can be elegantly arranged.

One also needs to incorporate electroweak symmetry breaking. The standard Higgs
mechanism is triggered by the breaking of supersymmetry. Supersymmetry
requires that are two Higgs doublets, of course.

\subsubsection{Supersymmetry Breaking, etc.}

Local models are incompatible with gravity-mediated SUSY breaking.
However, for {\it gauge-mediated SUSY breaking} both the visible and
the hidden sectors, as well as their mediation, can be described in
a single effective field theory decoupled from gravity. According to
most experts, such models are favored for suppressing
flavor-changing neutral currents. When the model is extended to
incorporate a global completion and bring gravity back in the story,
there will be a small admixture of gravity mediation. A proposal in
the recent literature is that gauge-mediated supersymmetry breaking
arises due to {\it stringy instantons} (Euclidean D3-branes that
wrap the 4-cycle $S$) on an auxiliary 7-brane that has been dubbed a
{\it Peccei-Quinn 7-brane}.

Beasley, Heckman, and Vafa \cite{Beasley:2008dc,Beasley:2008kw} have
also proposed interesting mechanisms by which local F-theory models
can address each of the following:

\begin{itemize}

\item R-parity conservation

\item Peccei--Quinn symmetry and axion properties

\item the $\mu$ and $\mu/B_{\mu}$ problems

\item the size of neutrino masses

\item proton decay

\item hierarchies of Yukawa couplings

\end{itemize}

There are many more details of this approach that I cannot go into
here, including singularity enhancement, classification of fluxes,
nonexistence of heterotic and M-theory duals, determination of
superpotentials, threshold corrections, dark matter, sparticle
spectrum, Higgsing the Peccei--Quinn symmetry, messenger fields,
etc. This subject is still in its infancy, but it shows great
promise. It will be very interesting to see what predictions can be
made before the experimental results pour in and whether they turn
out to be correct.

\section{Conclusion}

The understanding of string theory continues to progress on many
fronts at an impressive rate. Many challenges remain, and it will
undoubtedly require many decades to answer some of the deepest
questions. However, there is no reason to be pessimistic. The
subject is already proving useful for various purposes ranging from
fundamental mathematics to particle physics phenomenology and
cosmology. New experimental and observational results will be very
helpful in stimulating further progress.

\section*{Acknowledgment}

I am grateful to the Ettore Majorana Foundation and Centre for Scientific
Culture for its hospitality during this delightful school.
This work was supported in part by the U.S. Dept. of Energy under
Grant No. DE-FG03-92-ER40701.

\end{document}